\documentclass[twocolumn]{aastex631}

\DeclareUnicodeCharacter{02BC}{\-}
\shortauthors{Yan et al.}
\usepackage{textcomp, gensymb}
\usepackage{booktabs}
\usepackage{soul}
\usepackage{float}

\newcommand{\dotdeg}{\rlap{.}^\circ}

\begin{document}

\title{Multifrequency Very Long Baseline Interferometry Imaging of the Subparsec-scale Jet in the Sombrero Galaxy (M\,104)}

\author[0009-0003-6680-1628]{Xi Yan}
\affiliation{Shanghai Astronomical Observatory, Chinese Academy of Sciences, 80 Nandan Road, Shanghai 200030, People's Republic of China} 
\affiliation{School of Astronomy and Space Sciences, University of Chinese Academy of Sciences, 19A Yuquan Road, Beijing 100049, People's Republic of China}

\author[0000-0002-7692-7967]{Ru-Sen Lu}
\affiliation{Shanghai Astronomical Observatory, Chinese Academy of Sciences, 80 Nandan Road, Shanghai 200030, People's Republic of China}
\affiliation{Key Laboratory of Radio Astronomy and Technology, Chinese Academy of Sciences, A20 Datun Road, Chaoyang District, Beijing, 100101, People's Republic of China}
\affiliation{Max-Planck-Institut f\"ur Radioastronomie, Auf dem Hügel 69, D-53121 Bonn, Germany}

\correspondingauthor{Ru-Sen Lu}
\email{rslu@shao.ac.cn}

\author[0000-0001-7369-3539]{Wu Jiang}
\affiliation{Shanghai Astronomical Observatory, Chinese Academy of Sciences, 80 Nandan Road, Shanghai 200030, People's Republic of China}

\author[0000-0002-4892-9586]{Thomas P. Krichbaum}
\affiliation{Max-Planck-Institut f\"ur Radioastronomie, Auf dem Hügel 69, D-53121 Bonn, Germany}

\author[0000-0001-9969-2091]{Fu-Guo Xie}
\affiliation{Shanghai Astronomical Observatory, Chinese Academy of Sciences, 80 Nandan Road, Shanghai 200030, People's Republic of China}
\affiliation{Key Laboratory for Research in Galaxies and Cosmology, Shanghai Astronomical Observatory, Chinese Academy of Sciences, \\
80 Nandan Road, Shanghai 200030, People's Republic of China}

\author[0000-0003-3540-8746]{Zhi-Qiang Shen}
\affiliation{Shanghai Astronomical Observatory, Chinese Academy of Sciences, 80 Nandan Road, Shanghai 200030, People's Republic of China}
\affiliation{Key Laboratory of Radio Astronomy and Technology, Chinese Academy of Sciences, A20 Datun Road, Chaoyang District, Beijing, 100101, People's Republic of China}

\begin{abstract}
We report multi-frequency and multi-epoch VLBI studies of the sub-parsec jet in Sombrero galaxy (M\,104, NGC\,4594). Using Very Long Baseline Array data at 12, 22, 44, and 88\,GHz, we study the kinematics of the jet and the properties of the compact core. The sub-parsec jet is clearly detected at 12 and 22\,GHz, and the inner jet base is resolved down to $\sim70$ Schwarzschild radii ($R_{\rm s}$) at 44\,GHz. The proper motions of the jet are measured with apparent sub-relativistic speeds of $0.20\pm0.08\,c$ and $0.05\pm0.02\,c$ for the approaching and the receding jet, respectively. Based on the apparent speed and jet-to-counter-jet brightness ratio, we estimate the jet viewing angle to be larger than $\sim37^{\circ}$, and the intrinsic speed to be between $\sim0.10\,c$ and $0.40\,c$. Their joint probability distribution suggests the most probable values of the viewing angle and intrinsic speed to be ${66^{\circ}}^{+4^\circ}_{-6^\circ}$ and $0.19\pm0.04\,c$, respectively. We also find that the measured brightness temperatures of the core at 12, 22 and 44\,GHz are close to the equipartition brightness temperature, indicating that the energy density of the radiating particles is comparable to the energy density of the magnetic field in the sub-parsec jet region. Interestingly, the measured core size at 88\,GHz ($\sim25\pm5\,R_{s}$) deviates from the expected frequency dependence seen at lower frequencies. This may indicate a different origin for the millimeter emission, which can explained by an Advection Dominated Accretion Flow (ADAF) model. This model further predicts that at 230 and 340\,GHz, the ADAF may dominate the radio emission over the jet.
\end{abstract}
\keywords{galaxies: active --- galaxies: individual (M\,104) --- galaxies: nuclei --- radio continuum: galaxies}

\section{Introduction} \label{sec:Introduction}

\begin{deluxetable*}{cccccc}
\tablecaption{Summary of M\,104 Observations\label{tab:observation summary}}
\tablewidth{10pt}
\tablehead{\colhead{Epoch} & \colhead{Stations} & \colhead{Freq. (GHz)} & \colhead{Beam size (mas $\times$ mas, deg)} & \colhead{ $I_{\rm peak}$ (Jy beam$^{-1}$) } & \colhead{$I_{\rm rms}$ (Jy beam$^{-1}$)}} 
\decimalcolnumbers 
\startdata
2021/02/08 & VLBA, -NL  & 12 & $1.22\times0.455$, -2.11 & 0.0478 
& $9.0\times10^{-5}$   \\
&  & 22  & $0.791\times0.278$, -7.15 & 0.0412 
& 9.6 $\times10^{-5}$  \\
2021/05/10 & VLBA, PT* & 12 & $1.3\times0.499$, -4.77 & 0.0706 
& $7.0\times10^{-5}$  \\
&  & 22 & $0.853\times0.289$, -8.93 & 0.0559 
&$1.1\times10^{-4}$  \\
2021/08/03 & VLBA   & 12 & $1.27\times0.481$, -4.04 & 0.0644 
& $7.8\times10^{-5}$  \\
&  & 22 & $0.857\times0.263$, -11 & 0.0550 
& $1.9\times10^{-4}$  \\
2021/02/05  & VLBA, -HN,-NL,-PT  & 44 & $0.368\times0.145$, -2.42 & 0.0542 & $6.1\times10^{-4}$\\
& KP, LA, MK, OV & 88 & ... & ... & ... \\
\enddata
\tablecomments{
Column (1): Observing date.
Column (2): Participating stations. The minus sign indicates stations that were not involved due to weather conditions or technical problems. Note that LCP channels of PT has no data in the May observations.
Column (3): Observing frequency.
Column (4): Full Width at Half Maximum (FWHM) and position angle (P.A.) of the synthesized beam.
Columns (5)-(6): Peak intensity and the rms noise level.
}
\end{deluxetable*}

\vspace{-0.8cm}
Low-Luminosity Active Galactic Nuclei (LLAGNs) are very common in the local universe, accounting for over one-third of nearby galaxies \citep[e.g.,][]{ho_nuclear_2008}. They exhibit intrinsically low bolometric luminosities, typically characterized by log\,($L_{\rm bol}/L_{\rm Edd})\la -3$. This phenomenon of sub-Eddington luminosities is often attributed to the presence of a hot, geometrically thick, optically thin Advection Dominated Accretion Flow (ADAF) surrounding the central supermassive black holes (SMBHs), characterized by a low mass-accretion rate and low radiation efficiency~\citep{Narayan_1994ApJ...428L..13N,Narayan_1995ApJ...444..231N,yuan_2014ARA&A..52..529Y}. Beyond a certain radius, this hot flow is truncated by a standard cold and geometrically thin disk \citep[Shakura–Sunyaev Disk, SSD;][]{Shakura_1973A&A....24..337S,Gammie_1999ApJ...516..177G,Quataert_1999ApJ...525L..89Q}. In addition, the presence of a relativistic jet is anticipated, essential for explaining the observed radio \citep[e.g.,][]{wu_origin_2005,Liu_2013ApJ...764...17L} and X-ray emission \citep[e.g.,][]{Xie_2017ApJ...836..104X}. 
Such a three-component model has successfully interpreted the peculiar spectral energy distributions (SEDs) of LLAGNs \citep[e.g.,][]{Nemmen_2014MNRAS.438.2804N}.

Understanding the physical processes in these systems, especially in the immediate vicinity of the SMBHs, remains a challenge. This is mainly due to their complicated and apparently weak nuclear activity and their large distances from us. However, high-resolution very long baseline interferometry (VLBI) techniques at millimeter wavelengths have made it possible to probe directly the vicinity of some of these nearby SMBHs \citep[e.g.,][]{EHT_2019ApJ...875L...1E,Lu_2023}. Studying the properties of jets near the central SMBHs can provide valuable insights into the underlying physical processes in the nuclear regions.

M\,104, an early-type spiral galaxy (Sa), is classified as a low ionization nuclear emission region galaxy \citep{Heckman_1980A&A....87..152H,HO_1997ApJS..112..315H}. It is located in the Virgo cluster at a distance of 9.55$\pm$0.31\,Mpc \citep{mcquinn_distance_2016} and appears nearly edge-on with an inclination angle of $84\dotdeg3\pm0\dotdeg2$ \citep{rubin_rotation_1985,Sutter_2022ApJ...941...47S}.
Previous optical and near-infrared observations have revealed the presence of a SMBH at its center, with a mass of $\sim1\times10^{9} M_\odot$ \citep[e.g.,][]{Kormendy_1988ApJ...335...40K,Emsellem_1994A&A...285..739E,kormendy_1996ApJ...473L..91K,Menezes_2015ApJ...808...27M}. As a result, 1\,mas corresponds to a spatial scale of only $\sim$ 0.046\,pc or 484 $R_{\rm s}$. This makes M\,104 a prominent target for high-resolution VLBI studies, allowing us to probe the black hole environment down to a few tens of $R_{\rm s}$.

However, current VLBI results on the M\,104 jet are still very limited. Early mas-scale VLBI observations failed to detect extended jet emission due to sensitivity limitations \citep{Shaffer_1979ApJ...233L.105S,Graham_1981A&A....97..388G,Preston_1985AJ.....90.1599P}. Until 2013, \citet{hada_evidence_2013} revealed clearly two-sided jets extending along the northwest-southeast direction. They also measured core shifts and derived several physical parameters of the jet, such as the jet viewing angle and intrinsic speed. However, no further VLBI studies have been reported since then. In this paper, we present new multi-frequency and multi-epoch  Very Long Baseline Array (VLBA) observations of the M\,104 jet.

The paper is structured as follows. In Section\,\ref{sec:Observations and Data Reduction}, we describe the observations and data reduction. In Section\,\ref{sec:Results and Discussion}, the results and discussion are presented. Finally, a summary of the findings is provided in Section\,\ref{sec:summary}.

\section{Observations and Data Reduction} \label{sec:Observations and Data Reduction}
\subsection{12 and 22 GHz data} \label{subsec:12 and 22 GHz}
We performed quasi-simultaneous VLBA observations of M\,104 at 12 and 22\,GHz for three epochs in February, May, and August 2021 (Table\,\ref{tab:observation summary}). All data were recorded at a data rate of 4096\,Mbps using 2-bit quantization. The total bandwidth of 512\,MHz was split into four intermediate frequency (IF) bands in both right and left circular polarization (RCP and LCP). The quasars 3C279 and J1239-1023 served as fringe finder and phase calibrator, respectively. 

Following standard data reduction procedures, we performed amplitude and phase calibration using NRAO's Astronomical Image Processing System \citep[AIPS;][]{Greisen2003}. The residual phases, delays, and delay rates were calibrated using fringe fitting, which was performed in two steps. First, the constant single-band delays and phase offsets in different IFs were removed using high signal-to-noise ratio (S/N) calibrator scans. Then, a global fringe fitting was run to remove the residual single- and multi-band delays and to solve the fringe rates. The prior amplitude calibration, including opacity correction, was performed using station-based system temperature measurements, gain-elevation curves, and weather information. The bandpass correction was derived from the fringe finder scans by using the auto- and cross-correlation data. After the calibration, the data were averaged over all IFs and exported to DIFMAP \citep{Shepherd_Difmap_1997ASPC..125...77S} for imaging.

\subsection{44 and 88 GHz data} \label{subsec:44 and 88 GHz}
On 2021 Feb 5, we performed Source-Frequency-Phase-Referencing \citep[SFPR;][]{Rioja2011AJ....141..114R} observations of M\,104 with the VLBA at 44 and 88\,GHz (Table\,\ref{tab:observation summary}). The data were recorded at a data rate of 4096\,Mbps with 2-bit quantization. The bandwidth for both RCP and LCP was 512\,MHz, divided into 4 IFs. As mentioned in Section\,\ref{subsec:12 and 22 GHz}, 3C279 and J1239-1023 were observed as calibrators. Due to the short coherence time, a frequency switching cycle of 80\,s was chosen for both the calibrator and the target source. Within each cycle, 40\,s were spent at 44\,GHz, and the remaining 40\,s were spent at 88\,GHz. An average time of $\sim$10\,s per half-cycle was used to switch between the 44 and 88\,GHz feed horns, leading to a similar on-source integration time of $\sim$30\,s at each frequency. In addition, the frequency switching cycles on J1239-1023 ($1\dotdeg23$ apart on the sky) were interleaved every $\sim$12 minutes.

We analyzed the SFPRed data following the strategies specifically described in \citet{jiang_vlbi_2018}. Amplitude and bandpass calibration were performed in a similar manner as described in Section\,\ref{subsec:12 and 22 GHz}. The phase calibration was done in the following steps. First, we calibrated the ionospheric dispersive delays using the ionospheric model provided by the Jet Propulsion Laboratory. Then, the constant single-band delays and phase offsets at both 44 and 88\,GHz were removed. We performed global fringe fitting at 44\,GHz to solve for the single-band and multi-band residual delay, rate, and phase solutions. We multiplied the 44\,GHz residual phase solutions by the frequency ratio of 2 and applied frequency phase transfer (FPT) to the 88\,GHz data. However, due to the limited weather conditions during the observations, only the phase solutions from four antennas within a two-hour period at 44\,GHz were suitable for FPT (see Table\,\ref{tab:observation summary}). After calibration, the data were imported into DIFMAP for further analysis. Iterative phase and amplitude self-calibration was performed on the 44\,GHz data during imaging, while non-imaging analysis was performed on the 88\,GHz data.

\subsection{Flux density correction} 
\label{subsec: Flux density correction}
Our observations were affected by the flux density issues associated with the VLBA observing system. As a result, the total flux densities in our observations at all observing frequencies are generally lower than those reported in \citet{hada_evidence_2013}. At 12 and 22\,GHz, the exact origin of the flux density issue is not identified, but it can be due to systematic errors associated with the sampler modes\footnote{\href{https://science.nrao.edu/enews/14.4/index.shtml}{https://science.nrao.edu/enews/14.4/index.shtml\#vlba\_flux.}}. Based on the findings of the the MOJAVE team\footnote{Monitoring Of Jets in Active galactic nuclei with VLBA Experiments (MOJAVE), \href{https://www.cv.nrao.edu/MOJAVE/}{https://www.cv.nrao.edu/MOJAVE/}.} at 15\,GHz, we chose a scaling factor of $1.2\pm0.2$.

At 44 and 88\,GHz, our observations were primarily affected by pointing errors of the VLBA, as seen by the Blazar Research Group at Boston University\footnote{\href{https://www.bu.edu/blazars/}{https://www.bu.edu/blazars/}.} (BU group, A. Marscher, private comm.). To derive the scaling factors, we compared the amplitude of 3C\,279 in this study (epoch: 2021.02.08) with that provided by the BU group (epoch: 2021.01.22). The corresponding ratios of the flux densities (in units of Jy) are $13.5/26.1$ at 44\,GHz and $6.4/18.4$ at 88\,GHz. Therefore, we applied correction factors of $1.9\pm0.3$ and $2.9\pm0.5$ for M\,104 at 44 and 88\,GHz, respectively.

Note that the flux densities may still be inaccurate after this correction since pointing errors are source-dependent. Considering the previously reported total flux density of $91.2\pm9.1$\,mJy at 43\,GHz \citep{hada_evidence_2013}, we treated the scaled-up flux densities at 44 and 88\,GHz as upper limits (see Table\,\ref{tab:core properties}). During the data calibration and analysis, we chose not to scale the visibility amplitudes directly. Instead, we applied the scaling factors only to the flux densities of the components derived from model fitting reported in Tables \ref{tab:components} and \ref{tab:core properties}. It is worth to note that the scaling does not affect the measurements of the core sizes discussed in Section\,\ref{subsec:The frequency dependence of the core size}.

\begin{figure*}[htbp!]
\centering
    \includegraphics[width=0.75\textwidth]{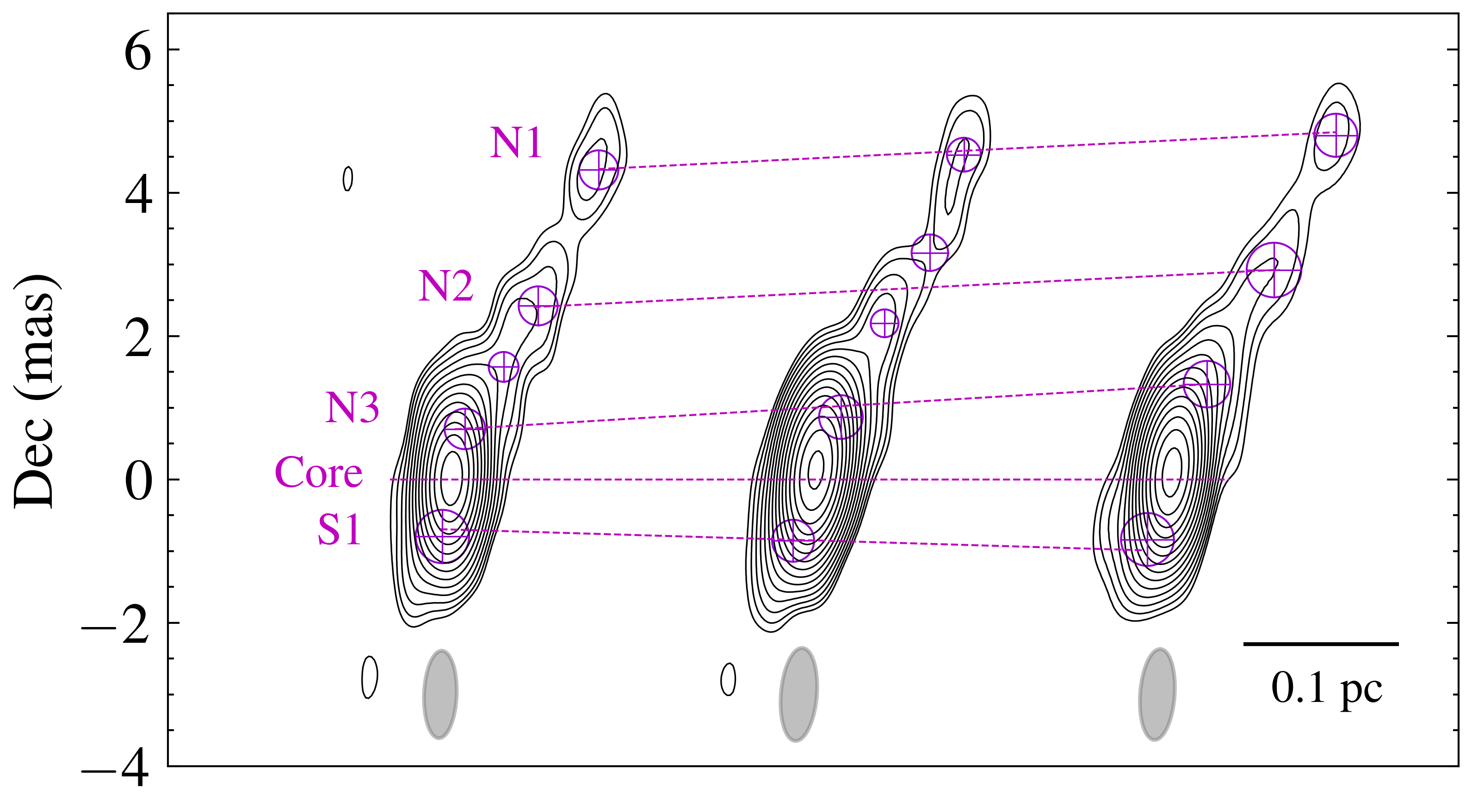}
    \quad
    \includegraphics[width=0.75\textwidth]{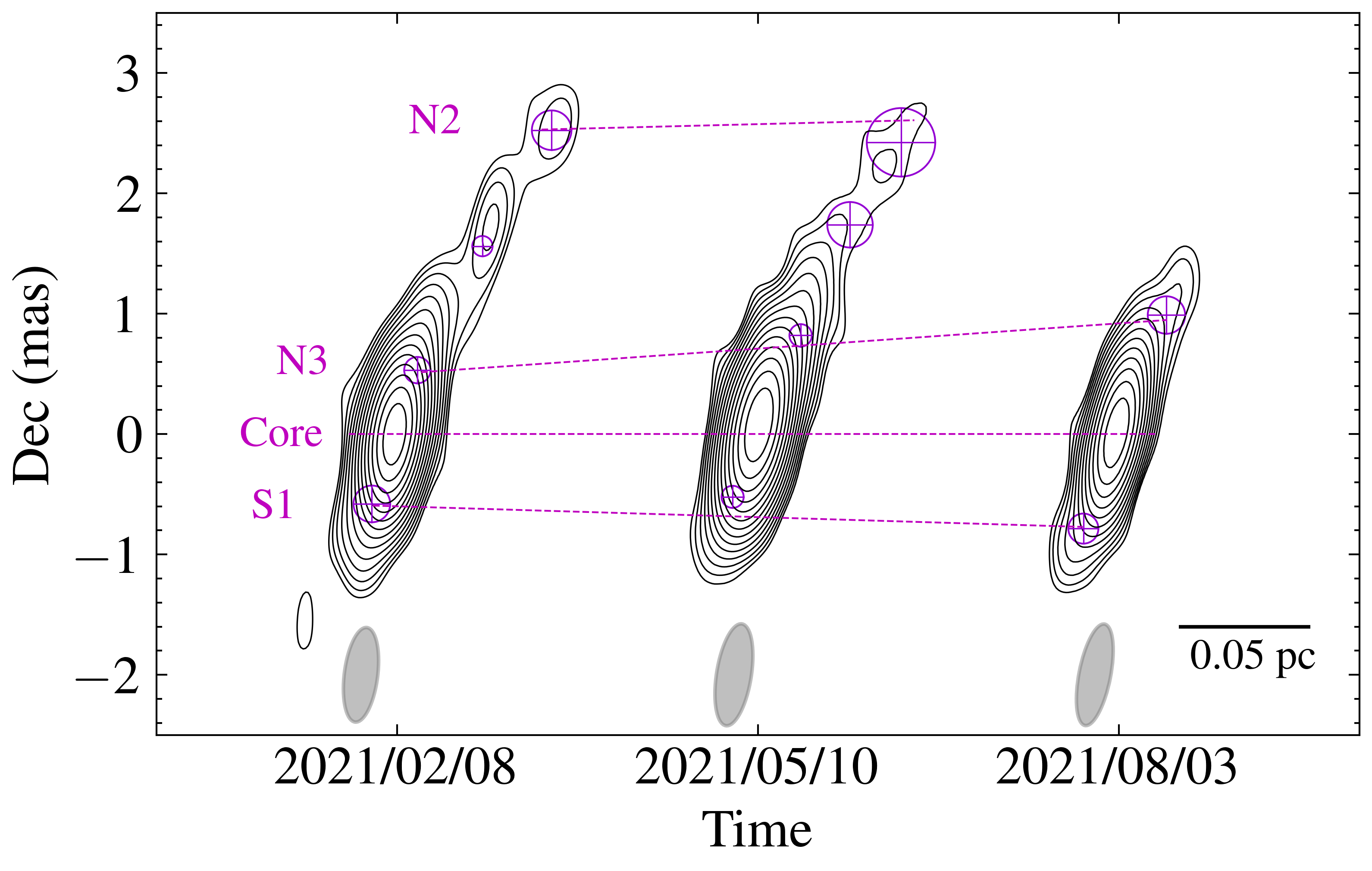}
\caption{Uniformly weighted CLEAN images of the M\,104 jet obtained with VLBA at 12 (top) and 22\,GHz (bottom) in 2021. The purple circles indicate the fitted circular Gaussian components, with labels showing on the left. The dashed lines represent the best-fit line of the proper motions. The light gray filled ellipses at the bottom of each panel represent the synthesized beam. Contours start from the 3 times rms noise level (see Table\,\ref{tab:observation summary}) and increase by a factor of $\sqrt{2}$.}
\label{fig:M104_12/22G_images}
\end{figure*}

\begin{figure*}[htbp!]
\begin{center}
    \includegraphics[width=0.44\textwidth]{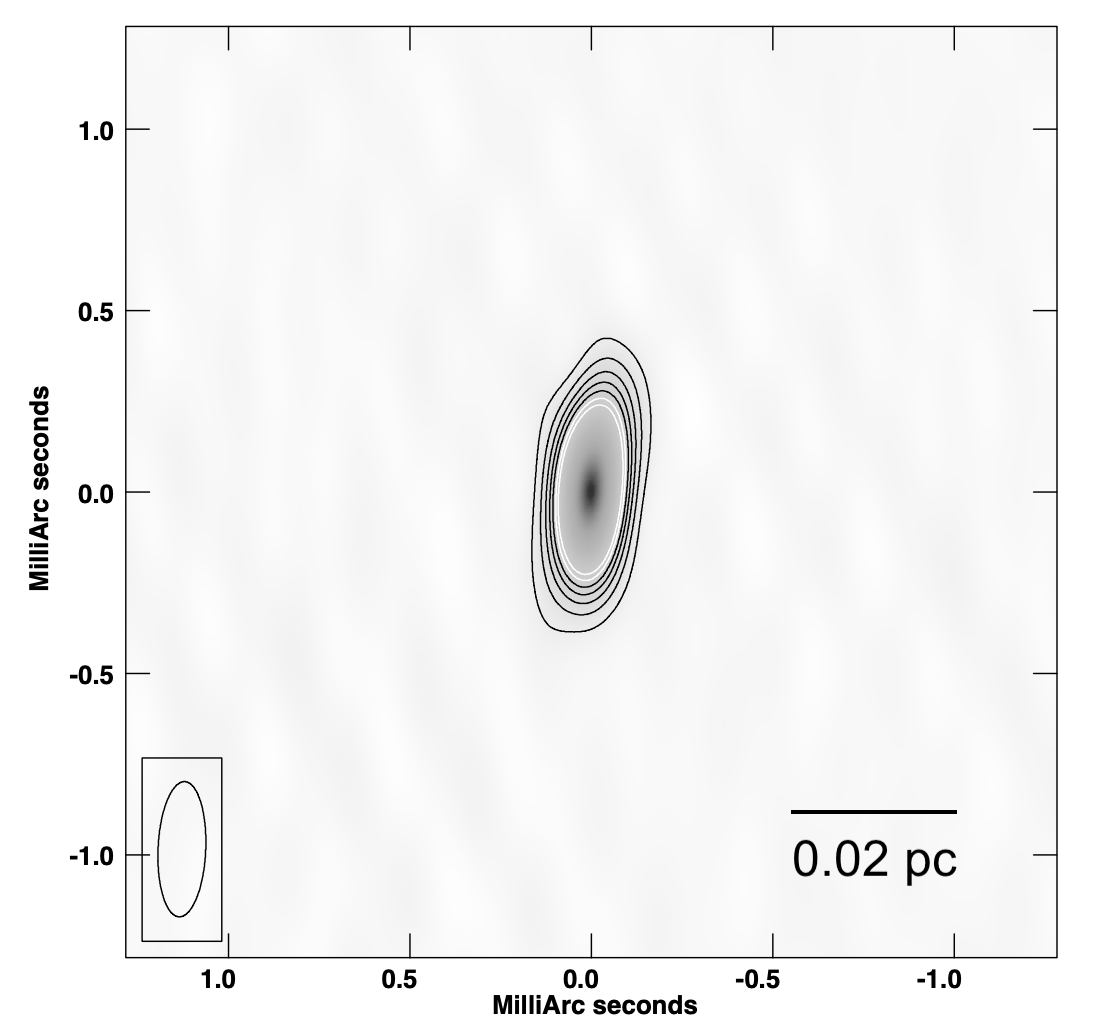}
    \includegraphics[width=0.52\textwidth]{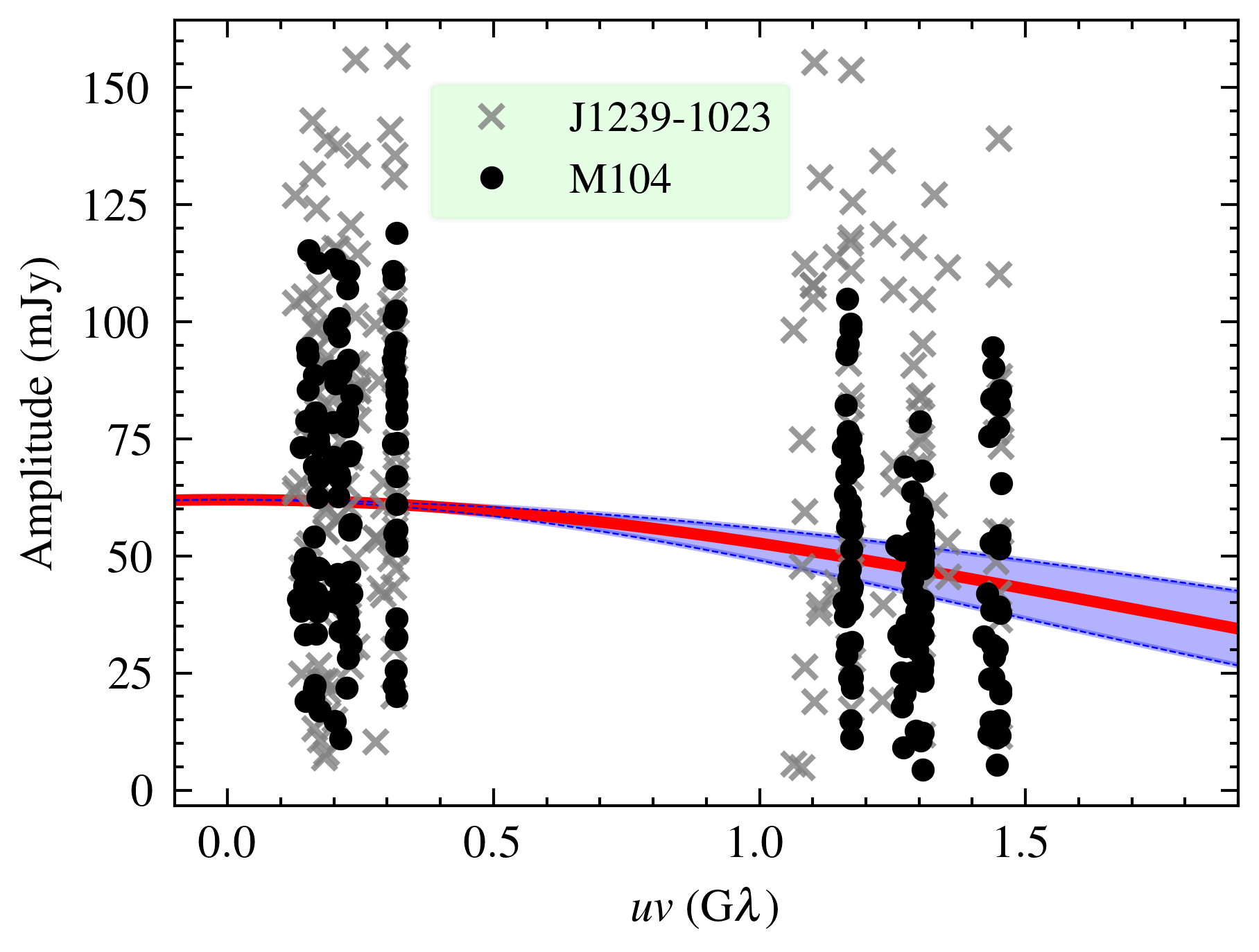}
    \caption{Left: Uniformly weighted CLEAN image of M\,104 nucleus observed at 44\,GHz. The synthesized beam is shown in the bottom left corner of the image. Contours start from the 3 times rms noise level (see Table\,\ref{tab:observation summary}) and increase by a factor of $\sqrt{2}$. Right: Visibility amplitude versus $uv$-distance plot of the 88\,GHz data. The solid red line denotes the fitted circular Gaussian model, while the shaded area indicates the associated error range (1$\sigma$ uncertainty). Also shown are the amplitude of the nearby compact calibrator J1239-1023.}
\label{fig:M104_44/88G_images}
\end{center}
\end{figure*}

\section{Results and Discussion} \label{sec:Results and Discussion}
\subsection{Source morphology} \label{subsec:Source morphology}
The CLEAN images at 12, 22, and 44\,GHz are shown in Figures\,\ref{fig:M104_12/22G_images} and \ref{fig:M104_44/88G_images}. Table\,\ref{tab:observation summary} summarizes the image parameters. We note that the last 22\,GHz map shows a lack of extended jet structure compared to the others. This can be attributed to the unfavorable weather conditions during the August observations, which resulted in a reduced number of fringe detection on several short baselines. However, the impact of these weather conditions on the 12\,GHz map was relatively small.

At 44\,GHz, the inner compact nuclear structure was clearly resolved down to a physical scale of $\sim$ 70 $R_{s}$ (Figure\,\ref{fig:M104_44/88G_images}, left). This is consistent with the phase-referenced result reported by \citet{hada_evidence_2013}. Unfortunately, imaging at 88\,GHz was difficult due to the very limited $u$-$v$ coverage. Therefore, we attempted to fit a model directly to the visibility data. We employed a circular Gaussian model to characterize the source structure, as it  provides a significant improvement in fitting (p-value $\ll$ 0.05) over a point-source model ($\delta$ function). The Gaussian fitting yielded a (lower-limit) flux density of $61.9\pm11.2$ mJy and a core size of $0.05\pm0.01$\,mas or $25\pm5\,R_{s}$ (see the right panel of Figure\,\ref{fig:M104_44/88G_images} and Table\,\ref{tab:core properties}). To validate the amplitude calibration, we show the amplitude as a function of {\it uv}-distance for the calibrator J1239-1023, which is consistent with a point-like morphology also suggested by previous observations of \citet{hada_evidence_2013} at 43\,GHz.

\subsection{Jet kinematics}\label{subsec:Jet kinematics in M104}
The sub-parsec structure detected at 12 and 22\,GHz allows us to study the kinematics of the M\,104 jet. To model the source structure, we used the model fitting method (i.e., the MODELFIT subroutine in DIFMAP) and screened out five circular Gaussian components: a core component, three forward jet components (N1, N2, and N3), and a counter jet component (S1). These components were cross-identified in different epochs and could be confirmed at different frequencies in the same epoch. The fitted results are shown in Table\,\ref{tab:components}.

To estimate the uncertainties of the fitting parameters of the component, we used the approximations described in \citet{Fomalont_1999ASPC..180..301F} and \citet{lee_2008AJ....136..159L}. However, this method is mainly based on the S/N and may underestimate the size and the position uncertainties \citep{hada_m87_2013ApJ...775...70H}. Therefore, for position errors less than 1/5 of the beam size along the jet direction, we used the latter as the error estimate. 

\begin{figure*}[htbp]
\begin{center}
\includegraphics[width=0.5\textwidth]{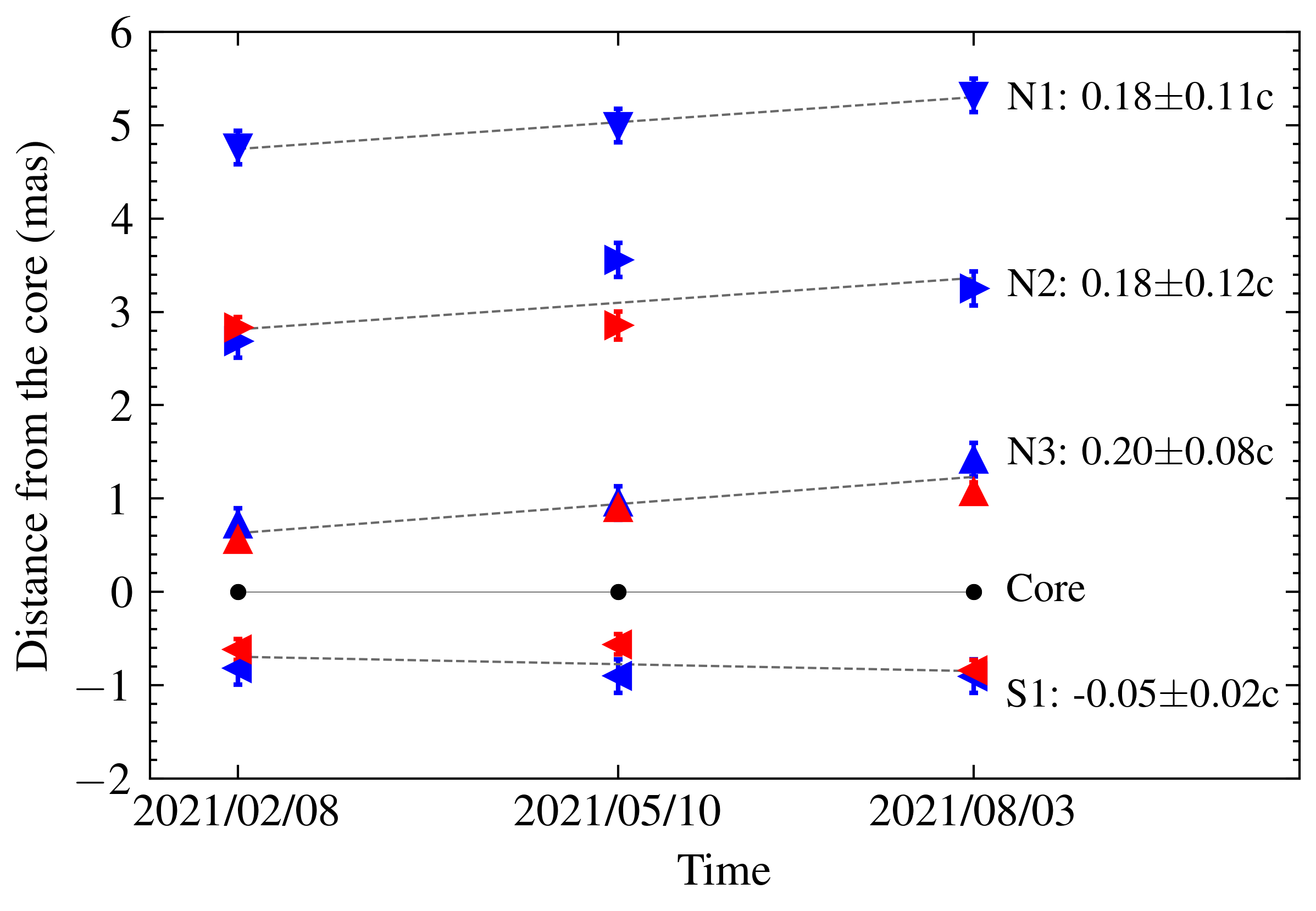}
\includegraphics[width=0.45\textwidth]{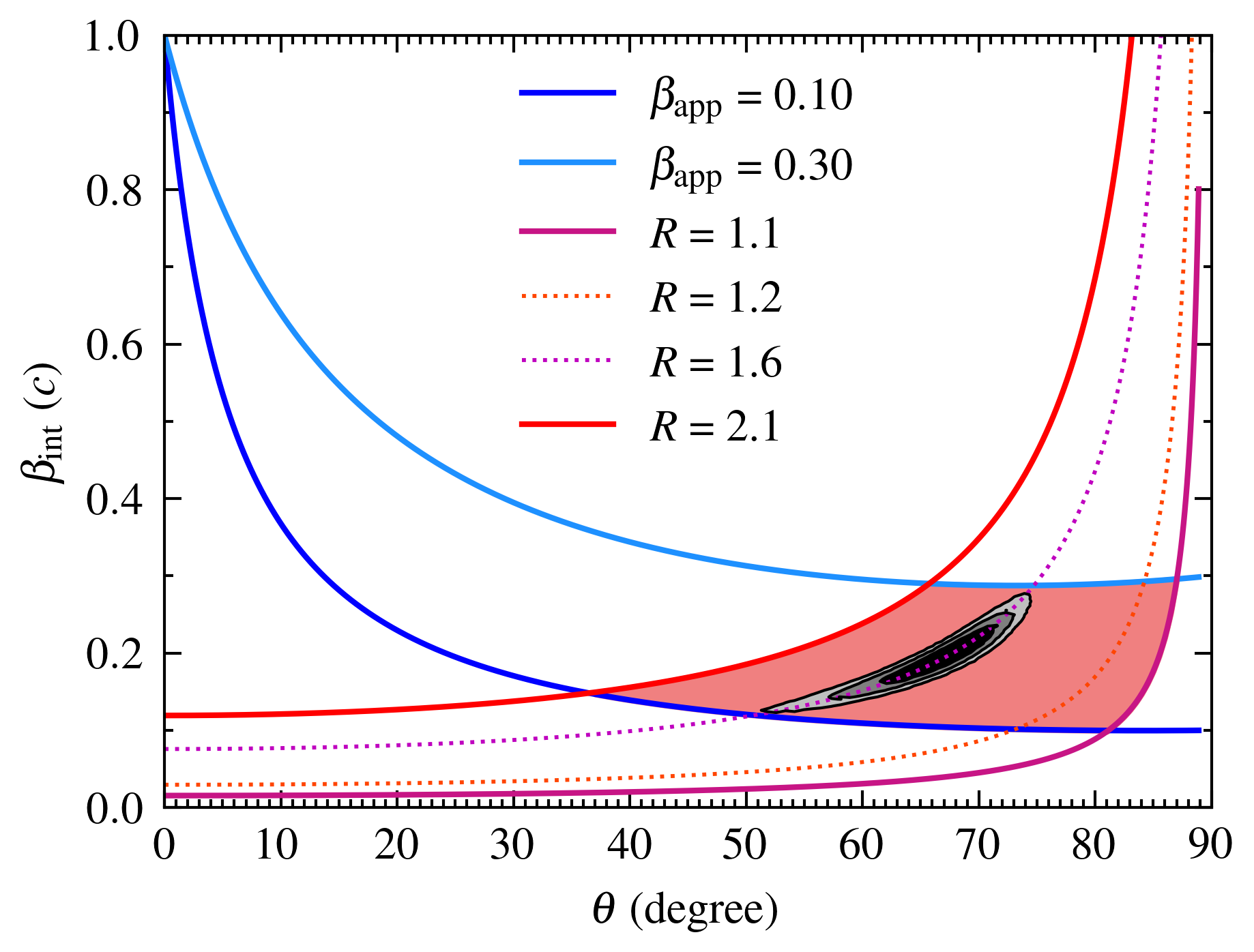}
\caption{Left: Core separation as a function of time for the cross-identified components. Different components are represented by different symbols. The components identified at 12 and 22\,GHz are shown in blue and red, respectively. Right: Possible range of the viewing angle and intrinsic velocity of M\,104 jet, derived from the observed apparent speed and the jet-to-counter-jet brightness ratio. The contours represent the joint probability distribution of the viewing angle and intrinsic speed, delineated at 1, 1.5, and 2 $\sigma$ equivalent levels.
\label{fig:M104_kinematics}}
\end{center}
\end{figure*}

Figure\,\ref{fig:M104_kinematics} (left) shows the measured proper motions of the M\,104 jet. Using linear fits to the radial distances from the core over time, we measured apparent speeds of $0.18\pm0.11\,c$, $0.18\pm0.12\,c$, and $0.20\pm0.08\,c$ for components N1, N2, and N3 in the approaching jet. In the counter-jet, we measured an apparent speed of $0.05\pm0.02\,c$ (commponent S1). For M\,104, this is the first direct observational evidence for proper motions on sub-parsec scales down to $\sim 0.25$\,pc.

\begin{deluxetable*}{ccccccccc}
\tabletypesize{\scriptsize}
\tablecaption{Properties of the Model-fitted Gaussian Components 
\label{tab:components}}
\tablehead{
\colhead{Epoch} & \colhead{$\chi^2_{\nu}$} & \colhead{ID} & \multicolumn{3}{c}{12\,GHz} & \multicolumn{3}{c}{22\,GHz} \\
\cmidrule(r){4-6}  \cmidrule(r){7-9} 
 & & & $S$ & $r$ & $d$ & $S$ & $r$ & $d$ \\
 & 12\,GHz/22\,GHz& & (mJy)  & (mas) & (mas) & (mJy)  & (mas) & (mas) \\
(1) & (2)& (3)& (4)& (5)& (6)& (7)& (8)& (9)}
\startdata
2021/02/08 & 1.03/0.99 & N1 &3.2$\pm$1.0 &4.76$\pm$0.18   &0.55$\pm$0.18 &$\lesssim0.8$ & ...& ... \\
 & & N2 & 3.8$\pm$1.1 &2.69$\pm$0.18  &0.54$\pm$0.18 &1.7$\pm$0.7 &2.83$\pm$0.11  &0.33$\pm$0.18 \\
 & & N3 &12.2$\pm$1.8 &0.72$\pm$0.18  &0.57$\pm$0.09 &5.6$\pm$1.1  &0.56$\pm$0.11  &0.22$\pm$0.04 \\
 & & S1 &9.6$\pm$1.7 &-0.82$\pm$0.18  &0.74$\pm$0.15 &4.6$\pm$1.0 &-0.62$\pm$0.11  &0.31$\pm$0.08 \\
2021/05/10 & 0.98/0.96 & N1 &1.7$\pm$0.6 &4.99$\pm$0.18   &0.47$\pm$0.18 &$\lesssim0.8$ & ...& ... \\
 & & N2 & 1.2$\pm$0.5 &3.56$\pm$0.18  &0.51$\pm$0.27 &1.9$\pm$0.8 &2.85$\pm$0.15  &0.57$\pm$0.30 \\
 & & N3 &8.9$\pm$1.4  &0.95$\pm$0.18  &0.61$\pm$0.10 &4.6$\pm$1.0  &0.89$\pm$0.11  &0.19$\pm$0.04 \\
 & & S1 &6.6$\pm$1.2 &-0.90$\pm$0.18  &0.59$\pm$0.11 &3.6$\pm$0.9 &-0.56$\pm$0.11  &0.18$\pm$0.04\\
2021/08/03 & 0.90/0.83 & N1 &2.5$\pm$0.9 &5.32$\pm$0.18   &0.60$\pm$0.22 &$\lesssim1.5$ & ...& ...\\
 & & N2 &3.4$\pm$1.0 &3.25$\pm$0.18  &0.76$\pm$0.26 & $\lesssim1.4$     &...  &...\\
 & & N3 &7.9$\pm$1.4  &1.42$\pm$0.18  &0.65$\pm$0.12 &4.1$\pm$1.4  &1.07$\pm$0.11  &0.31$\pm$0.11 \\
 & & S1 &8.4$\pm$1.5 &-0.91$\pm$0.18  &0.74$\pm$0.14 &4.9$\pm$1.4 &-0.84$\pm$0.11  &0.25$\pm$0.08   \\
\enddata
\tablecomments{
Column\,(1): Observing epoch. 
Column\,(2): The reduced $\chi^{2}$ of model fitting.
Column\,(3): Component label.
Column\,(4)-(6): Flux density, radial distance from the core, and size (FWHM) of the components at 12\,GHz.
Column\,(7)-(9): Same as Column\,(4)-(6), but at 22\,GHz.
}
\end{deluxetable*}

\subsection{Jet viewing angle}\label{subsec:The viewing angle of M104 jet}
The presence of the counter-jet allows us to measure the brightness ratio of the jet to the counter-jet.
In this study, we assume the two-sided jets in M\,104 to be intrinsically symmetric, with equal brightness and speed. Then using the determined brightness ratio and the apparent speed, we can constrain the viewing angle by the following equations:
\begin{equation} \label{eq:R}
\beta_{\rm int} = \frac{1}{\rm cos\theta}\left(\frac{R^{1/(2-\alpha)}-1}{R^{1/(2-\alpha)}+1}\right)
\end{equation}

\begin{equation} \label{eq:beta_int}
\beta_{\rm int} = \frac{\beta_{\rm app}}{\rm sin\theta+\beta_{\rm app}\rm cos\theta}
\end{equation}
where $R$ is the jet-to-counter-jet brightness ratio, $\beta_{\rm int}$ and $\beta_{\rm app}$ are the intrinsic speed and apparent speed, respectively, in units of $c$, $\theta$ is the jet viewing angle, and $\alpha$ is the spectral index ($S\propto\nu^{\alpha}$). We adopted $\alpha$=$-1.1$ based on the optically thin part of the observed jet by \citet{hada_evidence_2013}. 

To determine the jet-to-counter-jet brightness ratio, two rectangular boxes were placed on opposite sides of the jet in all 12 and 22\,GHz CLEAN images. These boxes are the same size and the same distance from the core. The averaged brightness ratio at 12 and 22\,GHz is about 1.2 and 1.1, respectively. These values are close to 1 and can be considered as lower limits. In addition to our measurements, we note that \citet{hada_evidence_2013} also measured the brightness ratio to be 1.6 at 15\,GHz and 2.1 at 24\,GHz. Therefore, by considering all these measurements, we adopted a brightness ratio range of 1.1 to 2.1.

By combining the brightness ratio with the measured apparent speed range ($0.10\,c\sim0.30\,c$), we found that the viewing angle of the M\,104 jet should be larger than $\sim37^{\circ}$, as shown in Figure\,\ref{fig:M104_kinematics} (right). We can also infer that the intrinsic speed of the M\,104 jet is between $\sim0.10\,c$ and $0.40\,c$. To further estimate the most probable values of the viewing angle and intrinsic speed, we employed Bayesian and Markov Chain Monte Carlo methods to derive the posterior joint probability distribution of $\theta$ and $\beta_{\rm int}$. The results are also shown in the the right panel of Figure\,\ref{fig:M104_kinematics}. The estimated most probable values for the viewing angle and intrinsic speed are ${66^{\circ}}^{+4^\circ}_{-6^\circ}$ and $0.19\pm0.04\,c$, respectively.

Notably, previous studies suggest a jet inclined closely to our line-of-sight. By comparing their measured core shifts with predictions from a compact jet model proposed by \citet{Falcke_1999A&A...342...49F}, \citet{hada_evidence_2013} indicated the viewing angle of the jet to be $\theta\la25^{\circ}$. We note that the model is developed based on a single conical jet shape and might not apply to the inner jet of M\,104. While the larger viewing angle range of $\theta\ga37^{\circ}$ reported in this study is supported by the VLBA detection of the almost symmetric two-sided jets at low frequencies. As can be seen from the Figure\,1 in \citet{hada_evidence_2013}, the images at 1.4, 2.3, 5.0, and 8.4\,GHz all show a prominent counter-jet that is equally extended as the approaching jet, suggesting that the viewing angle of M\,104 jet may not be too small. For comparison, the jet in M\,87 is believed to be inclined at about 20$^{\circ}$ \citep[i.e.,][]{mertens_kinematics_2016}, but it does not show a prominently visible counter-jet at these lower frequencies. Therefore, a larger viewing angle for the M\,104 jet may be more reasonable.

Based on the moderate amount of obscuring materials along the line of sight and their derived small viewing angle of the jet, \citet{hada_evidence_2013} suggested that the torus in M\,104 is nearly face-on, thus classifying it as a type I AGN. However, subsequent optical and near-infrared observations by \citet{Menezes_2013ApJ...765L..40M} and \citet{Menezes_2015ApJ...808...27M} detected a rotating dusty/molecular feature in M\,104 and interpreted it as a nearly edge-on torus. Assuming that the circumnuclear torus is perpendicular to the radio jets, our results also suggest an edge-on torus. Interestingly, we note that the nuclear emission even in many type II LLAGNs (i.e., with edge-on tori) can be relatively unobscured, as suggested by both X-ray and infrared observations~\citep[see detailed discussions in Sections 5.3-5.6 of][]{ho_nuclear_2008}. In addition, theoretical modeling also indicates that the broad-line region and the clumpy torus could become very weak and may even disappear under specific bolometric luminosity or Eddington ratio conditions~\citep[][]{ Nicastro_2000ApJ...530L..65N,Elitzur_2006ApJ...648L.101E}.

\subsection{Frequency dependence of the core size} 
\label{subsec:The frequency dependence of the core size}
We measured the size of the VLBI core at 12, 22, 44, and 88\,GHz (Table\,\ref{tab:core properties}) in order to study its frequency dependence. To cover a wider frequency range, archival VLBA data at 2.3 and 8.7\,GHz were included, although they were observed for different purposes. These data were observed in August 2017 and July 2018 respectively. The data have been properly calibrated and are available in the Astrogeo database\footnote{Astronomy and Geodesy Scientific Data Products,\\ \href{https://astrogeo.smce.nasa.gov/}{https://astrogeo.smce.nasa.gov/}.}. We determined the averaged core size to be $1.30\pm0.16$ mas at 2.3\,GHz and to be $0.31\pm0.03$ mas at 8.7\,GHz.

\begin{figure}[t!]
\centering
    \includegraphics[width=0.47\textwidth]{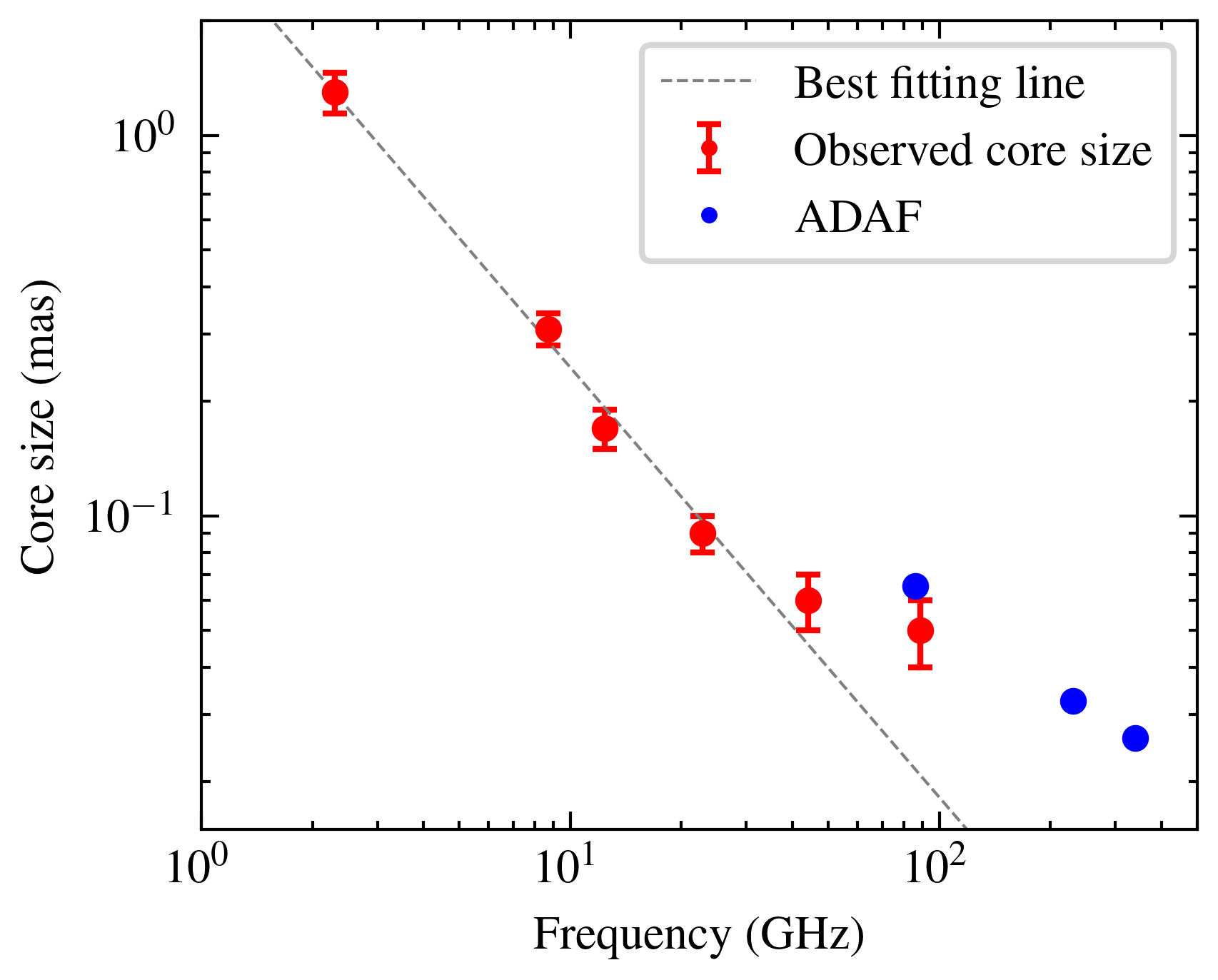}
    \caption{Frequency dependence of the VLBI core size. The measured core size and the predicted ADAF size are shown in red and blue, respectively. The grey dashed line represents the best power-law fit to the measured core sizes ($d\propto\nu^{-1.13\pm0.04}$). }
    \label{fig:core_size_frequency}
\end{figure}

In Figure\,\ref{fig:core_size_frequency}, we show the frequency dependence of the VLBI core size in M\,104. The dependence is best described by a power-law of $d\propto\nu^{-1.13\pm0.04}$ at lower frequencies, which is in full agreement with previous results \citep[$d \propto\nu^{-1.20\pm0.08}$ by][]{hada_evidence_2013}. At higher frequencies (i.e., 88\,GHz), however, the core size seems to deviate from this relationship. The flattening of this frequency-size dependence in the millimeter regime may indicate a different physical origin for the mm-VLBI core, which will be discussed in Section\,\ref{subsec:SED}. 

\begin{deluxetable*}{cccccc}
\tabletypesize{\scriptsize}
\tablecaption{Parameters of the Fitted Circular Guassian Core \label{tab:core properties}}
\tablewidth{0pt}
\tablehead{\colhead{Freq.\,(GHz)} & \colhead{Epoch}  & \colhead{$S$\,(mJy)} & \colhead{$d$\,(mas)} &\colhead{$T_{\rm B}$\,($\times 10^{10}\rm K$)} \\
 (1) & (2)& (3)& (4)& (5) 
}
\startdata
\hline
   & 2021/02/08  & $60.5\pm12.1$ &  $0.24\pm0.02$ & $0.8\pm0.2$\\
12 & 2021/05/10  & $49.4\pm9.9$ &  $0.13\pm0.01$ & $2.2\pm0.4$\\
   & 2021/08/03  & $54.2\pm10.8$ &  $0.14\pm0.01$ & $2.1\pm0.4$\\
Averaged &    & $54.7\pm10.9$ &  $0.17\pm0.02$ & $1.4\pm0.3$\\
\hline
   & 2021/02/08    & $50.2\pm10.0$&  $0.08\pm0.01$ & $2.0\pm0.4$\\ 
22 & 2021/05/10  & $46.4\pm9.3$ &  $0.11\pm0.01$ & $1.0\pm0.2$\\ 
   & 2021/08/03   & $65.6\pm13.1$&  $0.07\pm0.01$ & $3.4\pm0.7$\\ 
Averaged &     & $54.1\pm10.8$ &  $0.09\pm0.01$ & $1.7\pm0.3$\\
\hline
44 & 2021/02/05  & $62.3\pm 8.7 \lesssim S < 118.4\pm35.5$ &  $0.06\pm0.01$ & $1.1\pm0.2\lesssim T_{\rm B}<2.1\pm0.7$\\
88 & 2021/02/05  & $61.9\pm 11.2 \lesssim S <179.5\pm89.8$ &  $0.05\pm0.01$ & $0.4\pm0.1\lesssim T_{\rm B} <1.1\pm0.6$\\
\enddata
\tablecomments{
Column (1): Observing frequency. 
Column (2): Observing epoch. 
Columns (3)-(5): The flux density, size (FWHM), and brightness temperature of the core.}
\end{deluxetable*}

\subsection{Brightness temperature of the core} \label{subsec:The brightness temperature of the core}
Using the model parameters of the core listed in Table\,\ref{tab:core properties}, we derived its brightness temperature according to the equation 
\begin{equation}
T_{\rm B} = 1.22\times10^{12}\frac{S (1+z)}{\nu^{2}d^{2}} {\rm K}
\end{equation}
where $\nu$ is the observing frequency in GHz, $S$ is the flux density of the core in Jy, $d$ is the angular size in mas, and $z$=0.003416 is the redshift of M\,104. Considering the relatively low jet-to-counter-jet brightness ratio, we choose a Doppler factor of $\delta \approx 1$.

As is shown in Table\,\ref{tab:core properties}, the measured brightness temperatures at 12, 22, and 44\,GHz are basically close to the equipartition brightness temperature of $T_{\rm B,eq} \approx 5\times10^{10}\,\rm K$ \citep{readhead_1994ApJ...426...51R}. This suggests that the energy density of radiating particles is comparable to that of the magnetic field in the sub-parsec jet region. Furthermore, we also note that the brightness temperature at 88\,GHz is between $(0.4\pm0.1 \sim 1.1\pm0.6)\times10^{10}\,\rm K$, which possibly points to a highly magnetized jet base \citep[i.e.,][]{Lee_2016}.

\vspace{1cm}
\subsection{Spectra and broad-band SED modeling} \label{subsec:SED}
Using the model fitting results of the 12/22\,GHz data sets, we briefly examined the spectral properties of the core and the identified jet components. For the core, the averaged flux densities at 12 and 22\,GHz are quite close (see Table\,\ref{tab:core properties}), indicating a flat spectrum. Conversely, the components on both the approaching jet (N2 and N3) and the receding jet (S1) show a steep spectrum, with an averaged spectral index value of $\alpha = -0.97 \pm 0.07$. These findings align with the results obtained through a multi-frequency analysis by \citet{hada_evidence_2013} (see their Figures 4 and 5). Notably, the spectrum in the counter-jet exhibits an optically-thin synchrotron profile, indicating that free–free absorption (FFA) processes are not significant on the observing frequencies and the studied physical scales. For comparison, in sources like 3C\,84 \citep[e.g.,][]{Vermeulen_1994ApJ...430L..41V, Walker_1994ApJ...430L..45W}, Cygnus\,A \citep[e.g.,][]{Krichbaum_1998A&A...329..873K}, NGC\,1052 \citep[e.g.,][]{Kameno_2001PASJ...53..169K,Kadler_2004A&A...426..481K,Sawada-Satoh_2008ApJ...680..191S}, and NGC\,4261 \citep[e.g.,][]{Jones_1997ApJ...484..186J,Jones_2000ApJ...534..165J,Jones_2001ApJ...553..968J,Haga_2015ApJ...807...15H}, the counter-jet emission suffering from FFA by the accretion flow and/or torus exhibits an inverted steep spectrum with $\alpha>0$. 

\begin{figure}[htbp]
\begin{center}
\includegraphics[width=0.47\textwidth]{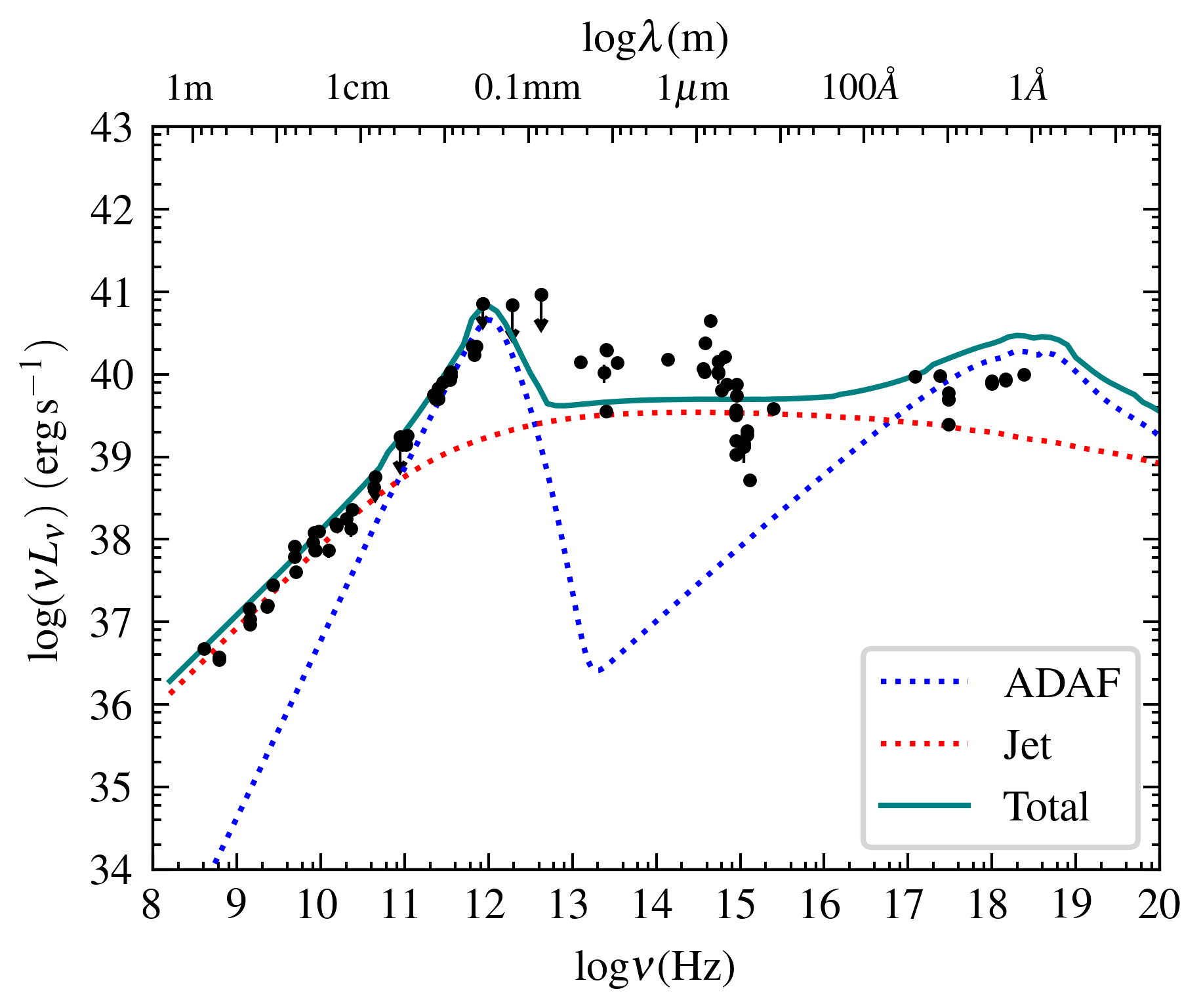}
\caption{The observed and modeled broad-band SED for M\,104. The dotted blue and red lines correspond to the emission from the ADAF and jet, respectively. The solid dark cyan line represents the sum of the emission from these two model components. 
\label{fig:M104_SED}}
\end{center}
\end{figure}

In Figure\,\ref{fig:M104_SED}, we show the broad-band SED of M\,104 by combining the high-resolution data observed at different wavelengths from literature \citep{Eracleous_2010ApJS..187..135E,Ontiveros_2023A&A...670A..22F}. Notably, multi-frequency radio observations reveal an emission ``excess" in the sub-millimeter spectral domain, i.e., the ``submillimeter bump"\footnote{We note that the thermal emission from the dust in the nucleus could potentially contaminate the observed ``submillimeter bump"~\citep{Sutter_2022ApJ...941...47S}.}.
To elucidate the origin of this bump, namely, whether it comes from the ADAF or the jet base, we applied a coupled ADAF-jet model to fit the broad-band SED of M\,104. This model has been successfully applied to many LLAGNs (e.g., \citealt{Quataert_1999ApJ...525L..89Q, wu_origin_2007, Nemmen_2014MNRAS.438.2804N, Xie_2016MNRAS.463.2287X}, and \citealt{yuan_2014ARA&A..52..529Y} for a review). The specific details of the model can be found in \citet{Xie_2016MNRAS.463.2287X}. For M\,104, the model parameters used here closely resemble those presented in \citet{Bandyopadhyay_2019MNRAS.490.4606B}.

As can be seen from Figure\,\ref{fig:M104_SED}, the radio emission in M\,104 is primarily jet-dominated at frequencies below about 100\,GHz. However, beyond this frequency range, the contribution from the accretion flow becomes increasingly significant. Consequently, the submillimeter bump might be ascribed to the synchrotron radiation from the thermal electrons in the ADAF. These results imply that the observed mm-VLBI core may be dominated by the emission from the accretion flow rather than the jet, which may explain the flattening of the frequency-size dependence in the millimeter regime in Figure\,\ref{fig:core_size_frequency}.

Therefore, we derived the expected ADAF sizes at 86, 230, and 340\,GHz using the model, which are also shown in Figure\,\ref{fig:core_size_frequency}. As can be seen, all the predicted sizes are significantly larger than the extrapolated core sizes from lower frequencies. Interestingly, the predicted ADAF size at 86\,GHz matches well with the measured core size at 88\,GHz, possibly suggesting the dominance of accretion flow over the jet. This is similar to the recent findings in M\,87, where 86\,GHz observations showed that the VLBI-core is spatially resolved into a ring-like structure, which is primarily dominated by the emission from the accretion flow \citep{Lu_2023}.

Finally, we remark that there is also a thermal bump component peaking at $\sim 1\mu$m in the SED of M\,104. As studied by \citet{Ontiveros_2023A&A...670A..22F}, this thermal bump can be adequately modeled with a cold, truncated SSD (see their Figure\,5). Thus, M\,104 serves as an exceptional target for a comprehensive high-resolution SED analysis, distinctly revealing three major components of LLAGN, including the radio jet, the hot ADAF, and the cold truncated SSD.

\section{Summary} \label{sec:summary}
In this study, we conducted multi-frequency and multi-epoch VLBA observations to study the jet kinematics and core properties in M\,104. Here, we summarize our conclusions:

\begin{enumerate}
\item We studied the jet kinematics and derived several physical properties for the jet: (1) The sub-parsec jet was clearly detected at 12 and 22\,GHz, and the inner jet was resolved down to $\sim$70 $R_{\rm s}$ at 44\,GHz. At 88\,GHz, we estimated the core size to be about $25\pm5\,R_{s}$. (2) The proper motions of the jet are measured with the apparent speeds of $0.20\pm0.08\,c$ and $0.05\pm0.02\,c$ for the approaching and the receding jet, respectively. (3) We estimated the jet viewing angle to be larger than ${37^{\circ}}$ and the intrinsic velocity to between $\sim 0.10\,c$ and $0.40\,c$, with the most probable values being ${66^{\circ}}^{+4^\circ}_{-6^\circ}$ and $0.19\pm0.04\,c$, respectively. The low apparent speed, intrinsic velocity, and the small jet-to-counter-jet brightness ratio indicate that the sub-parsec jet is sub-relativistic. (4) Both the jet and the counter-jet show an optically-thin synchrotron spectrum with an averaged spectral index value of $\alpha = -0.97 \pm 0.07$ at 12 and 22\,GHz, indicating that FFA processes are not significant on the studied sub-parsec scales.

\item To elucidate the origin of the submillimeter bump in M\,104, we applied a coupled ADAF-jet model to fit its broad-band SED. We found that the observed submillimeter bump may be ascribed to the synchrotron emission from the thermal electrons in the ADAF.

\item We investigated the frequency dependence of the VLBI core size over a frequency range between 2.3 and 88\,GHz. We found that the core size follows a power law relationship $d\propto\nu^{-1.13\pm0.04}$ at lower frequencies. However, at higher frequencies (i.e., 88\,GHz), the core size deviates from this power law, indicating a flattening of the frequency-size dependence in the millimeter regime. By combining with broad-band SED modeling, this flattening may indicate a different origin for the millimeter emission, which can be explained by an ADAF model. This model further predicts that at 230 and 340 GHz, the accretion flow may dominate the radio emission over the jet. Future high-resolution (sub)millimeter VLBI observations will be needed to further confirm this scenario.

\item We measured the brightness temperature of the core at various frequencies and found that they are close to the equipartition brightness temperature (except for 88\,GHz). This suggests that the energy density of the radiating particles is comparable to the energy density of the magnetic field in the sub-parsec jet region. Furthermore, the relatively low brightness temperature measured at 88\,GHz possibly indicates the presence of a highly magnetized jet base.

\end{enumerate}

We thank the anonymous referee for valuable comments and suggestions that improved the quality of the paper. We thank Dr.\,Jay Blanchard from NRAO for valuable discussions on flux density calibration of VLBA. We also thank Prof.\,Alan P. Marscher for his help on the flux calibration of the 44 and 88\,GHz data. We thank Prof.\,Alexander Pushkarev and Dr.\,Leonid Petrov for providing the calibrated data in the Astrogeo database, which enables us to show the results over a wide frequency range. The Astrogeo VLBI FITS image database is supported by Dr.\,Leonid Petrov and can be accessed via \href{http://astrogeo.smce.nasa.gov/vlbi_images}{http://astrogeo.smce.nasa.gov/vlbi$\_$images}. This work was supported by the National Science Fund for Distinguished Young Scholars of China (Grant No. 12325302), the Key Program of the National Natural Science Foundation of China (NSFC, grant No. 11933007), the Key Research Program of Frontier Sciences, CAS (grant No. ZDBS-LY-SLH011), the Shanghai Pilot Program for Basic Research, CAS, Shanghai Branch (JCYJ-SHFY-2021-013) and the Max Planck Partner Group of the MPG and the CAS. F.G.X. is supported in part by the National SKA Program of China (No. 2020SKA0110102), NSFC (grant No. 12373017, 12192220, and 12192223), and the Youth Innovation Promotion Association of CAS (Y202064). The Very Long Baseline Array is operated by the National Radio Astronomy Observatory, a facility of the National Science Foundation, operated under cooperative agreement by Associated Universities, Inc.

\bibliographystyle{aasjournal}
\bibliography{references}

\end{document}